\documentclass[showpacs,preprintnumbers,amsmath,amssymb]{revtex4}

\usepackage{graphicx}
\usepackage{dcolumn}
\usepackage{bm}

\begin{document}

\newcommand*{\PKU}{School of Physics and State Key Laboratory of Nuclear Physics and
Technology, Peking University, Beijing 100871,
China}\affiliation{\PKU}


\title{Vector meson $\omega$-$\phi$ mixing and their form factors in light-cone quark model}

\author{Wen Qian}\affiliation{\PKU}
\author{Bo-Qiang Ma}\email{mabq@phy.pku.edu.cn}\affiliation{\PKU}

\date{\today}

\begin{abstract}
The vector meson $\omega$-$\phi$ mixing is studied in two
alternative scenarios with different numbers of mixing angles, i.e.,
the one-mixing-angle scenario and the two-mixing-angle scenario, in
both the octect-singlet mixing scheme and the quark flavor mixing
scheme. Concerning the reproduction of experimental data and the
$Q^2$ behavior of transition form factors, one-mixing-angle scenario
in the quark flavor scheme performs better than that in the
octet-singlet scheme, while the two-mixing-angle scenario works well
for both mixing schemes. The difference between the two mixing
angles in the octet-singlet scheme is bigger than that in the quark
flavor scheme.
\end{abstract}



\maketitle

\section{\label{sec:level1}Introduction}

In the investigation of the internal structure of hadrons, quarks
and gluons are fundamental degrees of freedom whose behavior is
controlled by quantum chromodynamics (QCD). Because of  the
confinement property, perturbative QCD is only applicable at large
energy scale. To study hadronic properties at low energy scales,
nonperturbative effects must be taken into account. Some fundamental
nonperturbative QCD approaches are available, such as lattice QCD
methods and QCD sum rule techniques. Different relativistic quark
models also provide convenient ways to describe hadrons. The
light-cone constituent quark model, which is used as an effective
low-energy approximation to QCD, is one of them.

 The light-cone formalism~\cite{Lepage80,Brodsky82,Brodsky98} provides a convenient framework for
the relativistic description of hadrons in terms of quark and gluon
degrees of freedom. The hadronic wave function can be described by
light-cone Fock state expansion:\\
\begin{eqnarray}
|M\rangle &=& \sum |q\bar{q}\rangle \psi_{q\bar{q}}
        + \sum
        |q\bar{q}g\rangle \psi_{q\bar{q}g} + \cdots ,\\
|B\rangle &=& \sum |qqq\rangle \psi_{qqq}
        + \sum
        |qqqg\rangle \psi_{qqqg} + \cdots .
\end{eqnarray}
To simplify the problem, we take the minimal quark-antiquark Fock
state description of photons and mesons to calculate their
transition form factors, decay widths and other properties.

 The investigation of the electromagnetic transition processes
between pseudoscalar mesons and vector mesons is helpful to
understand the internal structure of mesons. The pseudoscalar
transition form factors $F_{\eta\gamma}(Q^2)$ and
$F_{\eta'\gamma}(Q^2)$ provide a good platform to study the $\eta$
and $\eta'$ mixing effects~\cite{Feldmann98,Cao99,Huang07}. There
are two mixing schemes when studying $\eta$-$\eta'$ mixing: the
octet-singlet mixing scheme and the quark flavor mixing scheme.
According to other works devoted to $\eta$-$\eta'$
mixing~\cite{Feldmann99,Feldmann00,Xiao05,Huang07}, both schemes
work well when only $\eta$ and $\eta'$ are involved. Sometimes a
second mixing angle is introduced to study $\eta$-$\eta'$ mixing,
especially when studying their decay
constants~\cite{Leutwyler98,Kaiser98,Feldmann98}. So there are two
alternative scenarios with different numbers of mixing angles: the
one-mixing-angle scenario and the two-mixing-angle scenario, in both
the octet-singlet mixing scheme and the quark flavor mixing scheme.

Similarly, $\omega$-$\phi$ mixing can be studied through transition
and decay processes. Naturally the $\omega$-$\phi$ mixing can also
be studied in two mixing schemes corresponding to the $\eta$-$\eta'$
mixing. Many works have been done concerning the $\omega$-$\phi$
mixing~\cite{Choi99,Escribano05,Feldmann00,Kucurkarslan06}, but only
in the one-mixing-angle scenario. In this paper we extend the
two-mixing-angle scenario into the study of the $\omega$-$\phi$
mixing.

   When studying the vector mesons, measurements of their
branching fractions and transition form factors provide important
tests of different models. The decays of $\omega$, $\phi$ have been
studied for many
years~\cite{Landsberg85,Dzhelyadin79,Dzhelyadin81,Viktorov80}. The
conversion decays $\phi\rightarrow\eta e^+e^-$ and
$\omega\rightarrow\pi e^+ e^-$ were collected with the CMD-2
detector in recent years~\cite{Achasov01,Akhmetshin05}, and not only
their branching fractions but also related transition form factors
$F_{\phi\rightarrow\eta\gamma^*}(Q^2)$,
$F_{\omega\rightarrow\pi\gamma^*}(Q^2)$ in the time-like region were
analysed. Recently there were also some new data about
$\omega\rightarrow\pi\gamma$ transition form factor extracted from
proton-proton collisions~\cite{Kaptari07}. With the light-cone
hadronic wave functions, the decay widths and transition form
factors of radiative decays $V\rightarrow P\gamma$ or $P\rightarrow
V\gamma$ (with $V=\omega,\phi$, $P=\pi,\eta,\eta'$) can be
calculated and compared with experimental data.
 In this
paper we try to study $\omega$-$\phi$ mixing using one-mixing-angle
scenario and two-mixing-angle scenario, respectively, with the
octet-singlet and the quark flavor mixing schemes in the light-cone
quark model. We give four sets of wave function parameters and
vector meson mixing angles of $\omega$-$\phi$ in different schemes
and compare the behaviors when predicting the $Q^2$ evolution of the
form factors.

In this paper, all the parameters of the model are re-determined by
the electroweak processes according to the constraints in previous
papers~\cite{Xiao02,Xiao03,Xiao05,Yu07} with new experimental data
from PDG~(2008)~\cite{PDG08}. In Sec.~\ref{sec:level2}, we give a
brief review of meson light-cone wave functions and form factor
calculations. In Sec.~\ref{sec:level3}, we exhibit the two mixing
angle scenarios in two mixing schemes in our calculation. In
Sec.~\ref{sec:level4}, numerical results of vector meson form factor
$Q^2$ evolution are presented and compared with experimental data.

\section{\label{sec:level2} Light-cone spin wave functions and transition form factors}
Based on light-cone quantization of
QCD~\cite{Brodsky82,Lepage80,Brodsky98}, the hadronic wave function
can be expressed using the Fock state expansion:
\begin{eqnarray}
|M (P^+, \mathbf{P}_\perp, S_z) \rangle
   =\sum_{n,\lambda_i}\int\prod_{i=1}^n \frac{\mathrm{d} x_i \mathrm{d}^2
        \mathbf{k}_{\perp i}}{\sqrt{x_i}~16\pi^3}
        16\pi^3\delta(1-\sum_{i=1}^n x_i)\delta^{(2)}(\sum_{i=1}^n \mathbf{k}_{\perp i})
               | n : x_i P^+, x_i \mathbf{P}_\perp+\mathbf{k}_{\perp i},
        \lambda_i \rangle
        \psi_{n/M}(x_i,\mathbf{k}_{\perp i},\lambda_i).
\end{eqnarray}
The wave function $\psi_{n/M}(x_i,\mathbf{k}_{\perp i},\lambda_i)$
is the amplitude for finding $n$ constituents with momenta $(x_i
P^+, \frac{m_i^2+(x_i \mathbf{P}_\perp+\mathbf{k}_{\perp i})^2}{x_i
P^+},x_i \mathbf{P}_\perp+\mathbf{k}_{\perp i})$, and $\lambda_i$ is
the helicity of the $i$-th constituent.

For simplicity we just take the minimal quark-antiquark Fock state
description of mesons to calculate their radii, decay widths,
transition form factors and other quantities. Thus a meson Fock
state ($n=2$) is described by,
\begin{eqnarray}
|M(P, S_Z)\rangle &=& \sum_{\lambda_1,\lambda_2}\int
\frac{\mathrm{d} x \mathrm{d}^2
        \mathbf{k}_{\perp}}{\sqrt{x(1-x)}16\pi^3}
           |x,\mathbf{k}_{\perp},
        \lambda_1,\lambda_2 \rangle
        \psi^{S_Z}_{M}(x,\mathbf{k}_{\perp},\lambda_1,\lambda_2)\\
&\doteq& \int \frac{\mathrm{d} x \mathrm{d}^2
        \mathbf{k}_{\perp}}{\sqrt{x(1-x)}16\pi^3}\Psi^{S_Z}_M(x,\mathbf{k}_\perp,\lambda_1,\lambda_2).
\end{eqnarray}
The model wave function is given by~\cite{Ji90,Ma93,Huang94}
\begin{eqnarray}
\Psi_{M}^{S_z}(x,\mathbf{k}_{\perp},\lambda_1,\lambda_2)
&=&\varphi(x,\mathbf{k}_{\perp})
\chi_M^{S_z}(x,\mathbf{k}_{\perp},\lambda_1,\lambda_2).
\end{eqnarray}
Since there is no explicit solution of the Bethe-Salpeter equation
for the mesons, harmonic oscillator wave function in the
Brodsky-Huang-Lepage(BHL) prescription~\cite{Brodsky82,Huang94} is
adopted to describe the quark momentum-space wave function,
\begin{eqnarray}
\varphi(x,\mathbf{k}_{\perp}) &=&
\varphi_{\mathrm{BHL}}(x,\mathbf{k}_{\perp})=A \exp
\left[-\frac{1}{8\beta^2}\left(\frac{m_1^2+\mathbf{k}_\perp^2}{x}
+\frac{m_2^2+\mathbf{k}_\perp^2}{1-x}\right)\right].
\end{eqnarray}
$\chi_M^{S_z}(x,\mathbf{k}_{\perp},\lambda_1,\lambda_2)$ is the spin
wave function which is obtained through the Melosh-Wigner rotation
or, equivalently, by proper vertices for mesons.

The instant-form state $\chi(T)$ and the front-form state $\chi(F)$
of spin-$\frac{1}{2}$ constituent quarks are related by the
Melosh-Wigner rotation~\cite{Melosh74,Kondratyuk80,Ma93}:
\begin{equation}
\left\{
\begin{array}{lll}
\chi_i^\uparrow(T) &=& w_i[(k_i^+ +m_i)\chi_i^\uparrow(F)-k_i^R
\chi_i^\downarrow(F)]\\
\chi_i^\downarrow(T)&=& w_i[(k_i^+ +m_i)\chi_i^\downarrow(F)+k_i^L
\chi_i^\uparrow(F)],
\end{array}
\right.
\end{equation}
where $w_i=1/\sqrt{2k_i^+ (k^0+m_i)}$, $k^{R,L}=k^1\pm k^2$,
$k^+=k^0+k^3=x \mathcal{M}$; here $k_i$ is the momentum of the quark
with mass $m_i$, the invariant mass of the composite system is
$\mathcal{M}=\sqrt{\frac{\mathbf{k_\perp}^2+m_1^2}{x}+\frac{\mathbf{k_\perp}^2+m_2^2}{1-x}}$.
The Melosh-Wigner rotation is essentially a relativistic effect due
to the transversal motions of quarks inside the hadrons, and such an
effect plays an important role in understanding the proton ``spin
puzzle" in the nucleon case~\cite{Ma91,Ma96}.

In the light-cone frame, momentums of the meson and its constituents
are:
\begin{eqnarray}
P &=& (P^+,P^-,\mathbf{P}_\perp)=(P^+,\frac{M^2}{P^+},\mathbf{0}_\perp),\\
k_1 &=&(x P^+, \frac{\mathbf{k}_\perp^2+m_1^2}{x
    P^+},\mathbf{k}_\perp),\\
k_2 &=&( (1-x) P^+, \frac{\mathbf{k}_\perp^2+m_2^2}{(1-x)
    P^+},-\mathbf{k}_\perp).
\end{eqnarray}
With these momentums substituted into the Melosh-Wigner rotation, we
get coefficients
$C^F_{M,S_z}(x,\mathbf{k}_\perp,\lambda_1,\lambda_2)$ in the spin
wave function
\begin{eqnarray}
\chi_M^{S_z}(x,\mathbf{k}_\perp,\lambda_1,\lambda_2)=\sum_{\lambda_1,
\lambda_2} C_{M,S_z}^F(x,\mathbf{k}_\perp,\lambda_1,\lambda_2)
\chi_1^{\lambda_1}(F)\chi_2^{\lambda_2}(F).
\end{eqnarray}

 The same wave function can be obtained if a proper vertex is chosen
for the meson~\cite{Choi97,Yu07}, that is,
\begin{equation}\label{wavefucntion}
\bar{u}(k_1,\lambda_1)\Gamma_{M}v(k_2,\lambda_2),
\end{equation}
with
\begin{equation}
\Gamma_P=\frac{1}{\sqrt{2} \sqrt{\mathcal{M}^2-(m_1-m_2)^2}}\gamma_5
\end{equation}
for pseudoscalar mesons, and
\begin{equation}
\Gamma_V=-\frac{1}{\sqrt{2}
\sqrt{\mathcal{M}^2-(m_1-m_2)^2}}(\gamma^\mu-\frac{k_1^\mu-k_2^\mu}{\mathcal{M}+m_1+m_2})\epsilon_\mu(P,S_z)
\end{equation}
for vector mesons.

The above two methods lead to the same meson light-cone spin wave
function:
\begin{eqnarray}
\chi_P(x,\mathbf{k}_\perp,\lambda_1, \lambda_2)=\sum_{\lambda_1,
\lambda_2} C^F_P(x,\mathbf{k}_\perp,\lambda_1,\lambda_2)
\chi_1^{\lambda_1}(F)\chi_2^{\lambda_2}(F)
\end{eqnarray}
for pseudoscalar mesons~\cite{Ma93,Huang94} (the subscription
$S_Z=0$ is omitted), where
\begin{equation}
\left\{
  \begin{array}{lll}
    C^F_P(x,\mathbf{k}_\perp,\uparrow,\uparrow)&=&\frac{1}{\sqrt{2}}w^{-1}(-k^L)(\mathcal{M}+m_1+m_2)\\
    C^F_P(x,\mathbf{k}_\perp,\uparrow,\downarrow)&=&\frac{1}{\sqrt{2}}w^{-1}((1-x)m_1+x m_2)(\mathcal{M}+m_1+m_2)\\
    C^F_P(x,\mathbf{k}_\perp,\downarrow,\uparrow)&=&\frac{1}{\sqrt{2}}w^{-1}(-(1-x)m_1-x m_2)(\mathcal{M}+m_1+m_2)\\
    C^F_P(x,\mathbf{k}_\perp,\downarrow,\downarrow)&=&\frac{1}{\sqrt{2}}w^{-1}(-k^{R})(\mathcal{M}+m_1+m_2),
  \end{array}
\right.
\end{equation}
with
$w=(\mathcal{M}+m_1+m_2)\sqrt{x(1-x)[\mathcal{M}^2-(m_1-m_2)^2]}$;
\begin{eqnarray}
\chi_V^{S_z}(x,\mathbf{k}_\perp,\lambda_1,
\lambda_2)=\sum_{\lambda_1,\lambda_2}
C_{V,S_z}^F(x,\mathbf{k}_\perp,\lambda_1,\lambda_2)
\chi_1^{\lambda_1}(F)\chi_2^{\lambda_2}(F)
\end{eqnarray}
for vector mesons~\cite{Yu07}, where
\begin{equation}
\left\{
  \begin{array}{lll}
    C_{V,1}^F(x,\mathbf{k}_\perp,\uparrow,\uparrow)&=&w^{-1}[\mathbf{k}_\perp^2+(\mathcal{M}+m_1+m_2)((1-x)m_1+x m_2)]\\
    C_{V,1}^F(x,\mathbf{k}_\perp,\uparrow,\downarrow)&=&w^{-1}[k^R(x\mathcal{M}+m_1)]\\
    C_{V,1}^F(x,\mathbf{k}_\perp,\downarrow,\uparrow)&=&w^{-1}[-k^R((1-x)\mathcal{M}+m_2)]\\
    C_{V,1}^F(x,\mathbf{k}_\perp,\downarrow,\downarrow)&=&w^{-1}[-(k^R)^2];
  \end{array}
\right.
\end{equation}
\begin{equation}
\left\{
  \begin{array}{lll}
    C_{V,0}^F(x,\mathbf{k}_\perp,\uparrow,\uparrow)&=&\frac{1}{\sqrt{2}}w^{-1}[k^L((1-2x)\mathcal{M}+(m_2-m_1))]\\
    C_{V,0}^F(x,\mathbf{k}_\perp,\uparrow,\downarrow)
            &=& \frac{1}{\sqrt{2}}w^{-1}[2 \mathbf{k}_\perp^2 + (\mathcal{M}+m_1+m_2)((1-x)m_1+x m_2)]\\
    C_{V,0}^F(x,\mathbf{k}_\perp,\downarrow,\uparrow)
            &=& \frac{1}{\sqrt{2}}w^{-1}[2 \mathbf{k}_\perp^2 + (\mathcal{M}+m_1+m_2)((1-x)m_1+x m_2)]\\
    C_{V,0}^F(x,\mathbf{k}_\perp,\downarrow,\downarrow)&=&\frac{1}{\sqrt{2}}w^{-1}[-k^R((1-2x)\mathcal{M}+(m_2-m_1))];
\end{array}
\right.
\end{equation}
\begin{equation}
\left\{
  \begin{array}{lll}
    C_{V,-1}^F(x,\mathbf{k}_\perp,\uparrow,\uparrow)&=&w^{-1}[-(k^L)^2]\\
    C_{V,-1}^F(x,\mathbf{k}_\perp,\uparrow,\downarrow)&=&w^{-1}[k^L((1-x)\mathcal{M}+m_2)]\\
    C_{V,-1}^F(x,\mathbf{k}_\perp,\downarrow,\uparrow)&=&w^{-1}[-k^L(x\mathcal{M}+m_1)]\\
    C_{V,-1}^F(x,\mathbf{k}_\perp,\downarrow,\downarrow)&=&w^{-1}[\mathbf{k}_\perp^2+(\mathcal{M}+m_1+m_2)((1-x)m_1+x
    m_2)].
\end{array}
\right.
\end{equation}
These coefficients satisfy the normalization condition
\begin{equation}
\sum_{\lambda_1,\lambda_2} C_{M,S_z}^{F \ast}( x,{\bf
k}_{\perp},\lambda_1,\lambda_2)C_{M,S_z}^F (x,{\bf
k}_{\perp},\lambda_1,\lambda_2)=1.
\end{equation}
Therefore, the Fock state expansion coefficients in the light-cone
wave function of the mesons are
\begin{equation}
\psi^{S_z}(x,{\bf k}_\perp,\lambda_1,\lambda_2)=C_{S_z}^F(x,{\bf
k}_\perp,\lambda_1,\lambda_2)\varphi_{\mathrm{BHL}}(x,{\bf k}).
\end{equation}

\begin{figure}[h]
\centering
\includegraphics[width=8cm,height=4cm]{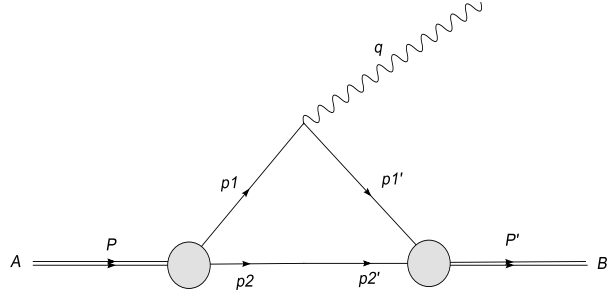}
\caption{\label{fig:diagram}The diagram for the transition form
factor $F_{A\rightarrow B\gamma}$ in the Drell-Yan-West frame.}
\end{figure}

 Pseudoscalar meson radii, the decay widths of pseudoscalar and vector mesons $P^\pm\rightarrow
\mu^\pm \nu$, $P^0\rightarrow \gamma\gamma$, $V\rightarrow e^+ e^-$,
$P\rightarrow V \gamma$, $V\rightarrow P \gamma$, and all the
transition form factors of these processes can be calculated in the
light-cone quark model using above meson wave functions. Supposing
that the instant-form wave functions of the mesons A and B in flavor
space are simply $|q_1 q_2\rangle$, the transition form factor of
$A\rightarrow B\gamma^*$ is defined by~\cite{Choi97}
\begin{eqnarray}
\langle B(P')|J^\mu |A(P,\lambda)\rangle
        =ieF_{A\rightarrow B\gamma}(Q^2)
        \varepsilon^{\mu\nu\rho\sigma}\epsilon_\nu(P,\lambda) P'_\rho
        P_\sigma,
\end{eqnarray}
where, $\epsilon(P,\lambda)$ is the polarization vector of the
vector meson. In the Drell-Yan-West~\cite{Drell70} frame, the
kinematics are, as shown in Fig.~\ref{fig:diagram},
\begin{equation}
\left\{
\begin{array}{lll}
q &=& (0,\frac{2P\cdot q}{P^+},\mathbf{q}_\perp)\\
P &=& (P^+,\frac{M_A^2}{P^+},\mathbf{0}_\perp)\\
P' &=& (P^+,\frac{M_B^2+\mathbf{q}_\perp^2}{P^+},-\mathbf{q}_\perp)\\
p_1 &=& (x P^+, \frac{m_1^2+\mathbf{k}_\perp^2}{x P^+},
    \mathbf{k}_\perp)\\
p_2 &=& ((1-x) P^+, \frac{m_2^2+\mathbf{k}_\perp^2}{(1-x) P^+},
    -\mathbf{k}_\perp)\\
p'_1 &=& (x P^+,
    \frac{m_1^2+(\mathbf{k}_\perp-\mathbf{q}_\perp)^2}{x P^+},
    x (-\mathbf{q}_\perp)+(\mathbf{k}_\perp-(1-x)\mathbf{q}_\perp))\\
p'_2 &=& ((1-x) P^+,
    \frac{m_2^2+\mathbf{k}_\perp^2}{(1-x) P^+},
    (1-x)(-\mathbf{q}_\perp)-(\mathbf{k}_\perp-(1-x)\mathbf{q}_\perp)).
\end{array}
\right.
\end{equation}
 Then, we get the transition form factor of $V\rightarrow
P\gamma^*$ or $P\rightarrow V\gamma^*$ calculated by the light-cone
quark model in the Drell-Yan-West frame:
\begin{eqnarray}
F_{A\rightarrow B\gamma^*} (Q^2) &=&\frac{\langle B(P')|J^+
|A(P,\lambda=+1)\rangle}
        {ie
        \varepsilon^{+\nu\rho\sigma}\epsilon_\nu(P,\lambda=+1) P'_\rho
        P_\sigma}\nonumber\\
&=& Q_{q_1} I_{VP\gamma}[m_1,m_2,A_A,\beta_A,A_B,\beta_B] - Q_{q_2}
I_{VP\gamma}[m_2,m_1,A_A,\beta_A,A_B,\beta_B],
\end{eqnarray}
in which
\begin{eqnarray}
I_{VP\gamma}[m_1,m_2,A_A,\beta_A,A_B,\beta_B] &=&2\int
\frac{\mathrm{d}x \mathrm{d}^2\mathbf{k}_\perp}{16\pi^3}
  \frac{1}{x(1-x)} \varphi^*_B(\mathbf{k}'_\perp)\varphi_A(\mathbf{k}_\perp)\nonumber\\
&&\times  \frac{((1-x)m_1+x
m_2)(\mathcal{M}+m_1+m_2)(1-x)+2(1-x)\mathbf{k}_\perp^2
\sin^2(\theta-\varphi)}
  {(\mathcal{M}+m_1+m_2)
  \sqrt{\mathcal{M}^2-(m_1-m_2)^2}\sqrt{\mathcal{M'}^2-(m_1-m_2)^2}}.
\end{eqnarray}
Here, $\varphi_{A,B}(\mathbf{k}_\perp,A_{A,B},\beta_{A,B})=A_{A,B}
\exp\left[-\frac{1}{8\beta^2_{A,B}}(\frac{m_1^2+\mathbf{k}_\perp^2}{x}+\frac{m_2^2+\mathbf{k}_\perp^2}{1-x})\right]$,
$\mathbf{k}'_\perp=\mathbf{k}_\perp-(1-x)\mathbf{q}_\perp$,
$\mathcal{M'}^2=\frac{m_1^2+\mathbf{k}^{'2}_\perp}{x}+\frac{m_2^2+\mathbf{k}^{'2}_\perp}{1-x}$;
$q$ is the momentum of the virtual photon, and in Drell-Yan-West
frame, $Q^2=-q^2=\mathbf{q}_\perp^2$; $Q_{q_1}$ and $Q_{q_2}$ are
electric charges of $q_1$ and $q_2$. The other formulas for decay
widths and form factors are presented in Appendix~\ref{app:a}.

\section{\label{sec:level3} Two mixing angle scenarios in two mixing schemes}
There are mainly two mixing schemes concerning $\eta$-$\eta'$ or
$\omega$-$\phi$ mixing. One is the octet-singlet mixing scheme
(denoted as 08)~\cite{Donoghue85,Gilman87},
\begin{eqnarray}
\left(\begin{array}{c}
         |\eta\rangle \\
         |\eta'\rangle
    \end{array}\right)
&=&\left(\begin{array}{cc}
     \cos \theta^S_{08} & -\sin \theta^S_{08} \\
     \sin \theta^S_{08} & \cos \theta^S_{08} \\
   \end{array}\right)
    \left(\begin{array}{c}
         |\eta_8\rangle \\
        |\eta_0\rangle \\
     \end{array}\right),\\
\left(\begin{array}{c}
         |\phi\rangle \\
        |\omega\rangle
    \end{array}\right)
&=&\left(\begin{array}{cc}
     \cos \theta^V_{08} & -\sin \theta^V_{08} \\
     \sin \theta^V_{08} & \cos \theta^V_{08} \\
   \end{array}\right)
\left(\begin{array}{c}
    |\omega_8\rangle \\
    |\omega_0\rangle \\
   \end{array}\right),
\end{eqnarray}
where, $\theta^S_{08}$ and $\theta^V_{08}$ are, respectively, the
pseudoscalar meson mixing angle and the vector meson mixing angle in
the octet-singlet mixing scheme. Here, the flavor SU(3) octet basis
is $|\psi_8\rangle = \frac{1}{\sqrt{6}}(u\bar{u}+d\bar{d}-2
s\bar{s})$ and singlet basis is $|\psi_0\rangle =
\frac{1}{\sqrt{3}}(u\bar{u}+d\bar{d}+ s\bar{s})$ (for $\psi=\eta$ or
$ \omega$). The other is quark-flavour basis mixing scheme (denoted
as qs)~\cite{Feldmann99,Cao99}:
\begin{eqnarray}
\left(\begin{array}{c}
    |\eta\rangle \\
    |\eta'\rangle
    \end{array}\right)
&=&\left(\begin{array}{cc}
        \cos \theta^S_{qs} & -\sin \theta^S_{qs} \\
        \sin \theta^S_{qs} & \cos \theta^S_{qs} \\
    \end{array}\right)
    \left(\begin{array}{c}
        |\eta_q\rangle \\
        |\eta_s\rangle \\
    \end{array}\right),\\
\left(\begin{array}{c}
    |\phi\rangle \\
    |\omega\rangle
    \end{array}\right)
&=&\left(\begin{array}{cc}
        \cos \theta^V_{qs} & -\sin \theta^V_{qs} \\
        \sin \theta^V_{qs} & \cos \theta^V_{qs} \\
    \end{array}\right)
    \left(\begin{array}{c}
        |\omega_q\rangle \\
        |\omega_s\rangle \\
    \end{array}\right),
\end{eqnarray}
where, $\theta^S_{qs}$ and $\theta^V_{qs}$ are the pseudoscalar
meson mixing angle and the vector meson mixing angle in the quark
flavor mixing scheme, and the quark flavor bases are $|\psi_q\rangle
= \frac{1}{\sqrt{2}}(u\bar{u}+d\bar{d})$, $|\psi_s\rangle =
s\bar{s}$ (for $\psi=\eta$ or $\omega$).

 The two schemes are equivalent to
each other by $\theta_{qs}=\theta_{08}+\arctan(\sqrt{2})$ when the
SU(3) symmetry is perfect. This relationship is not maintained when
we take into account the $SU(3)_f$ breaking by~\cite{Xiao05}:
\begin{eqnarray}
|\psi_8\rangle &=&
\frac{1}{\sqrt{6}}(u\bar{u}+d\bar{d})~\varphi_8^q(x,\mathbf{k}_\perp)
                -\frac{2}{\sqrt{6}}s\bar{s}~\varphi_8^s(x,\mathbf{k}_\perp), \\
|\psi_0\rangle &=&
\frac{1}{\sqrt{3}}(u\bar{u}+d\bar{d})~\varphi_0^q(x,\mathbf{k}_\perp)
                +\frac{1}{\sqrt{3}}s\bar{s}~\varphi_0^s(x,\mathbf{k}_\perp)
\end{eqnarray}
for the octet-singlet scheme, in which
\begin{equation}
\left\{
\begin{array}{lll}
\varphi_8^q(x,\mathbf{k}_\perp) &=& A_8
\exp[-\frac{m_q^2+\mathbf{k}_\perp^2} {8\beta_8^2 x(1-x)}]\\
\varphi_8^s(x,\mathbf{k}_\perp) &=& A_8
\exp[-\frac{m_s^2+\mathbf{k}_\perp^2} {8\beta_8^2 x(1-x)}]\\
\varphi_0^q(x,\mathbf{k}_\perp) &=& A_0
\exp[-\frac{m_q^2+\mathbf{k}_\perp^2} {8\beta_0^2 x(1-x)}]\\
\varphi_0^s(x,\mathbf{k}_\perp) &=& A_0
\exp[-\frac{m_s^2+\mathbf{k}_\perp^2} {8\beta_0^2 x(1-x)}],
\end{array}
\right.
\end{equation}
and
\begin{eqnarray}
|\psi_q\rangle &=&
\frac{1}{\sqrt{2}}(u\bar{u}+d\bar{d})~\varphi^q(x,\mathbf{k}_\perp),\\
|\psi_s\rangle &=& s\bar{s}~ \varphi^s(x,\mathbf{k}_\perp)
\end{eqnarray}
for the quark flavor scheme, in which,
\begin{equation}
\left\{
\begin{array}{lll}
 \varphi^q(x,\mathbf{k}_\perp) &=& A_q
\exp[-\frac{m_q^2+\mathbf{k}_\perp^2} {8\beta_q^2 x(1-x)}]\\
\varphi^s(x,\mathbf{k}_\perp) &=& A_s
\exp[-\frac{m_s^2+\mathbf{k}_\perp^2} {8\beta_s^2 x(1-x)}].
\end{array}
\right.
\end{equation}

In the octet-singlet mixing scheme, the decay constants of the
pseudoscalar mesons are given as follows,
\begin{eqnarray}
\left(
  \begin{array}{cc}
    f^8_\eta & f^0_\eta \\
    f^8_{\eta'} & f^0_{\eta'} \\
  \end{array}
\right) &=& \left(
  \begin{array}{cc}
    f_8 \cos{\theta^S_{08}} & -f_0 \sin{\theta^S_{08}} \\
    f_8 \sin{\theta^S_{08}} & f_0 \cos{\theta^S_{08}} \\
  \end{array}
\right).
\end{eqnarray}
The axial-vector anomaly and partial conservation of axial current
(PCAC) lead to~\cite{Escribano05}:
\begin{eqnarray}
\Gamma(\eta\rightarrow\gamma\gamma)
    &=&\frac{\alpha^2 m_\eta^3}{64\pi^3}
    \left(\frac{c_8}{f_{\eta_8}}\cos{\theta^S_{08}}-\frac{c_0}{f_{\eta_0}}\sin{\theta^S_{08}}
    \right)^2,\label{eq:Q2eta8}\\
\Gamma(\eta'\rightarrow\gamma\gamma)
    &=&\frac{\alpha^2 m_\eta^3}{64\pi^3}
    \left(\frac{c_8}{f_{\eta_8}}\sin{\theta^S_{08}}+\frac{c_0}{f_{\eta_0}}\cos{\theta^S_{08}}
    \right)^2.
\label{eq:Q2eta0}
\end{eqnarray}
Combining the above with
\begin{eqnarray}
F_{\eta\rightarrow\gamma\gamma^*}(Q^2) &=&
                F_{\eta_8\rightarrow\gamma\gamma^*}(Q^2)\cos{\theta^S_{08}}
                -F_{\eta_0\rightarrow\gamma\gamma^*}(Q^2)\sin{\theta^S_{08}},\\
F_{\eta'\rightarrow\gamma\gamma^*}(Q^2) &=&
                F_{\eta_8\rightarrow\gamma\gamma^*}(Q^2)\sin{\theta^S_{08}}
                +F_{\eta_0\rightarrow\gamma\gamma^*}(Q^2)\cos{\theta^S_{08}},
\end{eqnarray}
and their behavior when $Q^2\rightarrow\infty$,
\begin{eqnarray}
\lim_{Q^2\rightarrow\infty}Q^2
    F_{\eta_8\rightarrow\gamma\gamma^*}(Q^2)&=& c_8 f_{\eta_8},\label{eq:Q2infty1},\\
\lim_{Q^2\rightarrow\infty}Q^2
    F_{\eta_0\rightarrow\gamma\gamma^*}(Q^2) &=& c_0
    f_{\eta_0},\label{eq:Q2infty2},
\end{eqnarray}
one can constrain the $\eta$-$\eta'$ mixing angle and parameters,
while theoretical model calculation gives
\begin{eqnarray}
F_{\eta_8\rightarrow\gamma\gamma^*} (Q^2) &=&
\frac{1}{\sqrt{6}}(Q_u^2
    +Q_d^2)I_{P\gamma\gamma^*}[m_u,A_{\eta_8},\beta_{\eta_8}]-\frac{2}{\sqrt{6}} Q_s^2
    I_{P\gamma\gamma^*}[m_s,A_{\eta_8},\beta_{\eta_8}], \label{eq:eta82g},\\
F_{\eta_0\rightarrow\gamma\gamma^*} (Q^2) &=&
\frac{1}{\sqrt{3}}(Q_u^2
    +Q_d^2)I_{P\gamma\gamma^*}[m_u,A_{\eta_0},\beta_{\eta_0}]+ \frac{1}{\sqrt{3}}Q_s^2
    I_{P\gamma\gamma^*}[m_s,A_{\eta_0},\beta_{\eta_0}].
    \label{eq:eta02g}.
\end{eqnarray}

The decay constants and transition form factors of the vector mesons
$\omega$ and $\phi$ are

\begin{eqnarray}
\left(
  \begin{array}{c}
    f_\phi \\
    f_\omega\\
  \end{array}
\right) = \left(
  \begin{array}{cc}
    \cos\theta^V_{08} & -\sin\theta^V_{08} \\
    \sin\theta^V_{08} & \cos\theta^V_{08} \\
  \end{array}
\right) \left(
  \begin{array}{c}
    f_{\omega_8} \\
    f_{\omega_0} \\
  \end{array}
\right), \label{eq:fphi}
\end{eqnarray}

\begin{eqnarray}
\left(
  \begin{array}{c}
    F_{\phi\rightarrow\pi\gamma^*} (Q^2)\\
    F_{\omega\rightarrow\pi\gamma^*}(Q^2)\\
  \end{array}
\right) = \left(
  \begin{array}{cc}
    \cos\theta^V_{08} & -\sin\theta^V_{08} \\
    \sin\theta^V_{08} & \cos\theta^V_{08} \\
  \end{array}
\right) \left(
  \begin{array}{c}
    F_{\omega_8\rightarrow\pi\gamma^*}(Q^2) \\
    F_{\omega_0\rightarrow\pi\gamma^*}(Q^2) \\
  \end{array}
\right),
\end{eqnarray}

\begin{eqnarray}
\left(
  \begin{array}{c}
    F_{\phi\rightarrow\eta\gamma^*} (Q^2)\\
    F_{\phi\rightarrow\eta'\gamma^*} (Q^2)\\
    F_{\omega\rightarrow\eta\gamma^*}(Q^2)\\
    F_{\eta'\rightarrow\omega\gamma^*}(Q^2)
  \end{array}
\right) = \left(
  \begin{array}{cc}
    \cos\theta^V_{08} & -\sin\theta^V_{08} \\
    \sin\theta^V_{08} & \cos\theta^V_{08} \\
  \end{array}
\right) \otimes \left(
  \begin{array}{cc}
    \cos\theta^S_{08} & -\sin\theta^S_{08}\\
    \sin\theta^S_{08} & \cos\theta^S_{08} \\
  \end{array}
\right) \left(
  \begin{array}{c}
    F_{\omega_8\rightarrow\eta_8\gamma^*}(Q^2) \\
    F_{\omega_8\rightarrow\eta_0\gamma^*}(Q^2) \\
    F_{\omega_0\rightarrow\eta_8\gamma^*}(Q^2) \\
    F_{\omega_0\rightarrow\eta_0\gamma^*}(Q^2) \\
  \end{array}
\right),
\end{eqnarray}

in which,
\begin{equation}
\left\{
 \begin{array}{lll}
F_{\omega_8\rightarrow\pi\gamma^*}(Q^2) &=&
    \frac{1}{\sqrt{3}}I_{VP\gamma}[m_q,A_{\omega_8},\beta_{\omega_8},A_\pi,\beta_\pi]\\
F_{\omega_0\rightarrow\pi\gamma^*}(Q^2) &=&
    \frac{2}{\sqrt{6}}I_{VP\gamma}[m_q,A_{\omega_8},\beta_{\omega_8},A_\pi,\beta_\pi]\\
F_{\omega_8\rightarrow\eta_8\gamma^*}(Q^2) &=&
    \frac{1}{6}(\frac{2}{3}I_{VP\gamma}[m_q,A_{\omega_8},\beta_{\omega_8},A_{\eta_8},\beta_{\eta_8}]
    -\frac{8}{3}I_{VP\gamma}[m_s,A_{\omega_8},\beta_{\omega_8},A_{\eta_8},\beta_{\eta_8}])\\
F_{\omega_8\rightarrow\eta_0\gamma^*}(Q^2) &=&
    \frac{1}{\sqrt{18}}(\frac{2}{3}I_{VP\gamma}[m_q,A_{\omega_8},\beta_{\omega_8},A_{\eta_0},\beta_{\eta_0}]
    +\frac{4}{3}I_{VP\gamma}[m_s,A_{\omega_8},\beta_{\omega_8},A_{\eta_0},\beta_{\eta_0}])\\
F_{\omega_0\rightarrow\eta_8\gamma^*}(Q^2) &=&
    \frac{1}{\sqrt{18}}(\frac{2}{3}I_{VP\gamma}[m_q,A_{\omega_0},\beta_{\omega_0},A_{\eta_8},\beta_{\eta_8}]
    +\frac{4}{3}I_{VP\gamma}[m_s,A_{\omega_0},\beta_{\omega_0},A_{\eta_8},\beta_{\eta_8}])\\
F_{\omega_0\rightarrow\eta_0\gamma^*}(Q^2) &=&
    \frac{1}{3}(\frac{2}{3}I_{VP\gamma}[m_q,A_{\omega_0},\beta_{\omega_0},A_{\eta_0},\beta_{\eta_0}]
    -\frac{2}{3}I_{VP\gamma}[m_s,A_{\omega_0},\beta_{\omega_0},A_{\eta_0},\beta_{\eta_0}]).\label{eq:Fomega0eta0}
\end{array}
\right.
\end{equation}

In the quark flavor mixing scheme, the formulas are similar to those
in the octet-singlet scheme as shown in Appendix~\ref{app:b}.

Up to now we just use one-mixing-angle scenario in both the
octet-singlet and the quark flavor mixing scheme.
 We can also introduce two-mixing-angle scenario to do phenomenological
investigation, especially when studying the decay constants of
pseudoscalar mesons~\cite{Feldmann98,Feldmann98b}.

As stated in Ref.~\cite{Feldmann98b}, the Fock state decomposition
of a charge neutral meson can be generally expressed as:
\begin{eqnarray}
|M\rangle &=& C_M^8 |\psi_8\rangle + C_M^0 |\psi_0\rangle
    + C_M^g|gg\rangle + C_M^c|c\bar{c}\rangle + \cdots .
\end{eqnarray}
By truncating only the valence Fock states and doing
phenomenological analysis, two mixing angles could be introduced for
the meson state mixing. The relations are analogous to the mixing of
the pseudoscalar meson decay constants~\cite{Feldmann98}. To
simplify the problem we just assume that the mixing angles in the
valence Fock state decomposition are equal to those in the
pseudoscalar meson decay constant mixing.

Take the octet-singlet mixing scheme for example,
\begin{eqnarray}
\left(\begin{array}{c}
         |\eta\rangle \\
         |\eta'\rangle
    \end{array}\right)
&=&\left(\begin{array}{cc}
     \cos \theta^S_{8} & -\sin \theta^S_{0} \\
     \sin \theta^S_{8} & \cos \theta^S_{0} \\
   \end{array}\right)
    \left(\begin{array}{c}
         |\eta_8\rangle \\
        |\eta_0\rangle \\
     \end{array}\right),
\label{eq:etamixing}
\end{eqnarray}
where $\theta^S_{8},\theta^S_{0}$ are the two mixing angles
introduced for pseudoscalar mesons $\eta$-$\eta'$ in the
octet-singlet mixing scheme. Then the decay constants of the
pseudoscalar mesons are given by
\begin{eqnarray}
\left(
  \begin{array}{cc}
    f^8_\eta & f^0_\eta \\
    f^8_{\eta'} & f^0_{\eta'} \\
  \end{array}
\right) &=&
\left(
  \begin{array}{cc}
    f_8 \cos{\theta^S_8} & -f_0 \sin{\theta^S_0} \\
    f_8 \sin{\theta^S_8} & f_0 \cos{\theta^S_0} \\
  \end{array}
\right).
\end{eqnarray}
The axial-vector anomaly and PCAC lead to
\begin{eqnarray}
\Gamma(\eta\rightarrow\gamma\gamma^*)
    &=&\frac{\alpha^2 m_\eta^3}{64\pi^3}
    \left(\frac{\frac{c_8}{f_{\eta_8}}\cos{\theta^S_0}-\frac{c_0}{f_{\eta_0}}\sin{\theta^S_8}}
                        {\cos(\theta^S_0-\theta^S_8)}\right)^2,\label{eq:PCAC2angle1}\\
\Gamma(\eta'\rightarrow\gamma\gamma^*)
    &=&\frac{\alpha^2 m_\eta^3}{64\pi^3}
    \left(\frac{\frac{c_8}{f_{\eta_8}}\sin{\theta^S_0}+\frac{c_0}{f_{\eta_0}}\cos{\theta^S_8}}
                        {\cos(\theta^S_0-\theta^S_8)}\right)^2.
\end{eqnarray}
Combined with
\begin{eqnarray}
F_{\eta\rightarrow\gamma\gamma^*}(Q^2) &=&
                F_{\eta_8\rightarrow\gamma\gamma^*}(Q^2)\cos{\theta^S_8}
                -F_{\eta_0\rightarrow\gamma\gamma^*}(Q^2)\sin{\theta^S_0},\\
F_{\eta'\rightarrow\gamma\gamma^*}(Q^2) &=&
                F_{\eta_8\rightarrow\gamma\gamma^*}(Q^2)\sin{\theta^S_8}
                +F_{\eta_0\rightarrow\gamma\gamma^*}(Q^2)\cos{\theta^S_0},\label{eq:etap2g}
\end{eqnarray}
and their $Q^2\rightarrow\infty$ behavior in
Eqs.~(\ref{eq:Q2infty1},\ref{eq:Q2infty2}), theoretical formulas
Eqs.~(\ref{eq:eta82g},\ref{eq:eta02g}) can be used to constrain the
$\eta\eta'$ parameters.

Similarly, $\omega$-$\phi$ mixing can also be studied with
two-mixing-angle scenario:
\begin{eqnarray}
\left(\begin{array}{c}
         |\phi\rangle \\
        |\omega\rangle
    \end{array}\right)
&=&\left(\begin{array}{cc}
     \cos \theta^V_{0} & -\sin \theta^V_{8} \\
     \sin \theta^V_{0} & \cos \theta^V_{8} \\
   \end{array}\right)
\left(\begin{array}{c}
    |\omega_8\rangle \\
    |\omega_0\rangle \\
   \end{array}\right).
\label{eq:omegamixing}
\end{eqnarray}
In two-mixing-angle scenario in the octet-singlet mixing scheme, the
decay constants and transition form factors of the vector mesons
$\omega$ and $\phi$ are:

\begin{eqnarray}
\left(
  \begin{array}{c}
    f_\phi \\
    f_\omega\\
  \end{array}
\right) = \left(
  \begin{array}{cc}
    \cos\theta^V_8 & -\sin\theta^V_0 \\
    \sin\theta^V_8 & \cos\theta^V_0 \\
  \end{array}
\right) \left(
  \begin{array}{c}
    f_{\omega_8} \\
    f_{\omega_0} \\
  \end{array}
\right),
\end{eqnarray}

\begin{eqnarray}
\left(
  \begin{array}{c}
    F_{\phi\rightarrow\pi\gamma^*} (Q^2)\\
    F_{\omega\rightarrow\pi\gamma^*}(Q^2)\\
  \end{array}
\right) = \left(
  \begin{array}{cc}
    \cos\theta^V_8 & -\sin\theta^V_0 \\
    \sin\theta^V_8 & \cos\theta^V_0 \\
  \end{array}
\right) \left(
  \begin{array}{c}
    F_{\omega_8\rightarrow\pi\gamma^*}(Q^2) \\
    F_{\omega_0\rightarrow\pi\gamma^*}(Q^2) \\
  \end{array}
\right), \label{eqn:0-8 Vpig}
\end{eqnarray}

\begin{eqnarray}
\left(
  \begin{array}{c}
    F_{\phi\rightarrow\eta\gamma^*} (Q^2)\\
    F_{\phi\rightarrow\eta'\gamma^*} (Q^2)\\
    F_{\omega\rightarrow\eta\gamma^*}(Q^2)\\
    F_{\eta'\rightarrow\omega\gamma^*}(Q^2)
  \end{array}
\right) = \left(
  \begin{array}{cc}
    \cos\theta^V_8 & -\sin\theta^V_0 \\
    \sin\theta^V_8 & \cos\theta^V_0 \\
  \end{array}
\right) \otimes \left(
  \begin{array}{cc}
    \cos\theta^S_8 & -\sin\theta^S_0 \\
    \sin\theta^S_8 & \cos\theta^S_0 \\
  \end{array}
\right) \left(
  \begin{array}{c}
    F_{\omega_8\rightarrow\eta_8\gamma^*}(Q^2) \\
    F_{\omega_8\rightarrow\eta_0\gamma^*}(Q^2) \\
    F_{\omega_0\rightarrow\eta_8\gamma^*}(Q^2) \\
    F_{\omega_0\rightarrow\eta_0\gamma^*}(Q^2) \\
  \end{array}
\right). \label{eqn:0-8 Vetag}
\end{eqnarray}
When taking $\theta^S_8=\theta^S_0=\theta^S_{08}$ and
$\theta^V_8=\theta^V_0=\theta^V_{08}$, one returns back to the
one-mixing-angle scenario.

The two-mixing-angle scenario in the quark flavor mixing scheme is
similar to that in the above octet-singlet scheme, and it can be
obtained just by replacing the octet  bases with the quark flavor
bases as shown in Appendix~\ref{app:b}.

When the two mixing angles are not equal to each other, the mixing
matrices in Eqs.~(\ref{eq:etamixing}) and (\ref{eq:omegamixing}) are
not unitary. Also, due to the contributions from gluons, $c\bar{c}$,
and other higher Fock states, it is possible that the left valence
decomposition of the two mesons are not orthogonal to each other.
This justifies the two-mixing-angle scenario as a phenomenological
method to analyze the contributions from the valence part of
pseudoscalar and vector mesons.

In principle, the mixing angles in the valence Fock state
decomposition might be not the same as those in the pseudoscalar
meson decay constant mixing. Therefore one might introduce more
complicated scenarios of three mixing angles or even four mixing
angles, also with different combinations. However, such procedures
would be too complicated and the physical significance is also
obscure; hence we do not consider these complications further in our
work.

\section{\label{sec:level4}Numerical Results and predictions}
\subsection{\label{sec:level4.1}Set $\pi$, $K$, $\rho$ parameters}
Following Refs.~\cite{Xiao03,Xiao02,Xiao05,Yu07}, we use the decay
constants, radii and decay widths to determine $m_u=m_d=m_q$
(suppose the isospin symmetry), $m_s$, $A_M$, and $\beta_M$ (with
$M=\pi$, $K$, $\rho$). The experimental data are updated from PDG
(2008). The parameters and reproduced quantities of the mesons we
obtain are listed in Table~\ref{tab:table1} and
Table~\ref{tab:table2}.

\begin{table}
\caption{\label{tab:table1}Decay constants, charge radii and decay
widths of pseudoscalar and vector mesons for fitting $\pi$, $K$,
$\rho$ parameters. The experimental data are taken from PDG
(2008)~\cite{PDG08}.}
\begin{ruledtabular}
\begin{tabular}{cccc}
    & $F_{\mathrm{exp}}/f_{\mathrm{exp}}$~(GeV)~(Input)      & $F_{\mathrm{th}}/f_{\mathrm{th}}$~(GeV)~(Output) \\
\hline
    $f_{\pi^+}$  & $0.0922\pm0.0001$                      &  $0.0922$ \\
    $\langle r_\pi^2 \rangle$ $\mathrm{fm^2}$   & $0.45\pm0.01$   &  $0.45$\\
    $F_{\pi^0\rightarrow\gamma\gamma^*}(0)$ & $0.274\pm0.010$      &  $0.274$\\
    $f_{K^+}(K^+\rightarrow \mu\nu)$  & $0.1100\pm0.0006$   &  $0.1100$ \\
    $\langle r_{K^+}^2 \rangle$ $\mathrm{fm^2}$   & $0.31\pm0.03$ &  $0.31$\\
    $\langle r_{K^0}^2 \rangle$ $\mathrm{fm^2}$ & $-0.077\pm0.010$      & $-0.077$\\
    $f_{\rho}(\rho\rightarrow e^+ e^-)$  & $0.1564\pm0.0007$             &  $0.1564$ \\
    $F_{\rho^+\rightarrow\pi^+\gamma}(0)$ & $0.83\pm0.06$ & $0.83$\\
\end{tabular}
\end{ruledtabular}
\end{table}

\begin{table}
\caption{\label{tab:table2}Optimized parameters we get according to
the properties of the mesons in Table~\ref{tab:table1}.}
\begin{ruledtabular}
\begin{tabular}{c cccc ccccc}
 $m_u$  &$m_s$
        & $A_\pi$  &  $\beta_\pi$
        & $A_K$ & $\beta_K$
        & $A_\rho$ & $\beta_\rho$ \\
\hline
 $0.198 ~\mathrm{GeV}$ & $0.556~\mathrm{GeV}$
        & $47.36 ~\mathrm{GeV}^{-1}$ & $0.411 ~\mathrm{GeV}$
        & $68.73 ~\mathrm{GeV}^{-1}$ & $0.405 ~\mathrm{GeV}$
        & $48.585 ~\mathrm{GeV}^{-1}$ & $0.373 ~\mathrm{GeV}$
\end{tabular}
\end{ruledtabular}
\end{table}

\subsection{\label{sec:level4.2}Set $\eta\eta'$, $\phi\omega$
parameters in the one-mixing-angle scenario in two mixing schemes}
Take the octect-singlet scheme first. We accept the mixing angle of
$\eta$-$\eta'$ determined by taking into account the
$Q^2\rightarrow\infty$ behavior of the form factors of $\eta$,
$\eta'$~\cite{Xiao05,Cao99,Lepage80}, i.e., combining
Eqs.~(\ref{eq:Q2eta8}-\ref{eq:eta02g}) with experimental pole
formula, the pseudoscalar meson mixing angle can be solved:
\begin{eqnarray}
\tan \theta^S=
\frac{-(1+c^2)(\rho_1+\rho_2)+\sqrt{(1+c^2)^2(\rho_1+\rho_2)^2+4(c^2-\rho_1\rho_2)(1-c^2\rho_1\rho_2)}}
{2(c^2-\rho_1\rho_2)}, \label{eq:mixingangleS}
\end{eqnarray}
where
$\rho_1=\sqrt{\frac{\Gamma_{\eta\rightarrow\gamma\gamma^*}m_{\eta'}^3}
{\Gamma_{\eta'\rightarrow\gamma\gamma^*}m_{\eta}^3}}$,
$\rho_2=\frac{F_{\eta\rightarrow\gamma\gamma^*}(Q^2\rightarrow\infty)}
{F_{\eta'\rightarrow\gamma\gamma^*}(Q^2\rightarrow\infty)} =\left.
    \frac{F_{\eta\rightarrow\gamma\gamma^*}(0)\frac{1}{1+Q^2/\Lambda_\eta^2}}
   {F_{\eta'\rightarrow\gamma\gamma^*}(0)\frac{1}{1+Q^2/\Lambda_{\eta'}^2}}
   \right|_{Q^2\rightarrow\infty}
= \rho_1\frac{\Lambda_\eta^2}{\Lambda_{\eta'}^2}$. The pole-mass
parameters are taken as the CLEO Collaboration
results~\cite{CLEO98}:
\begin{eqnarray}
\Lambda_\eta=774 \pm 11 \pm 16 \pm 22 ~\mathrm{MeV},\,\,
    \Lambda_\eta=859 \pm 9\pm 18 \pm 20 ~\mathrm{MeV}.
\end{eqnarray}
In the octet-singlet mixing scheme the constants are
\begin{eqnarray}
c&=&\frac{c_0}{c_8},\\
c_P&=&(c_\pi,c_8,c_0)=(1,\frac{1}{\sqrt{3}},\frac{2\sqrt{2}}{\sqrt{3}}).
\end{eqnarray}
Thus the $\eta$-$\eta'$ mixing angle in octet-singlet scheme is
$\theta^S_{08}=-16.05^\circ$. Then we use the following constraints
to set the parameters of $\eta$ and $\eta'$:
\begin{eqnarray}
F_{\eta\rightarrow\gamma\gamma^*}(0)&=&\sqrt{\frac{4}{\alpha^2\pi
    M_\eta^3}\Gamma_{\eta\rightarrow\gamma\gamma}} ,
\end{eqnarray}
\begin{eqnarray}
F_{\eta'\rightarrow\gamma\gamma^*}(0)&=&\sqrt{\frac{4}{\alpha^2\pi
    M_{\eta'}^3}\Gamma_{\eta'\rightarrow\gamma\gamma}} ,
\end{eqnarray}
\begin{eqnarray}
F_{\rho\rightarrow\eta\gamma^*}(0)=\sqrt{\frac{3\Gamma_{\rho\rightarrow\eta\gamma}}{\alpha}
\left(\frac{2 m_\rho}{m_\rho^2-m_\eta^2}\right)^3},
\end{eqnarray}
\begin{eqnarray}
F_{\eta'\rightarrow\rho\gamma^*}(0)=\sqrt{\frac{\Gamma_{\eta'\rightarrow\rho\gamma}}{\alpha}
\left(\frac{2 m_{\eta'}}{m_{\eta'}^2-m_\rho^2}\right)^3}.
\end{eqnarray}

 The values of these constrains coming form experimental data are
displayed in the first column of Table~\ref{tab:table3}. With values
of $m_u$, $m_s$, $A_\rho$, $\beta_\rho$ set in
Sec.~\ref{sec:level4.1}, we can proceed to determine parameters of
$\eta$ and $\eta'$ by these constraints. The reproduced decay widths
given by theoretical fit with optimized parameters are displayed in
the second column of  Table~\ref{tab:table3}.

\begin{figure}[h]
\centering
\includegraphics[0,0][300,250]{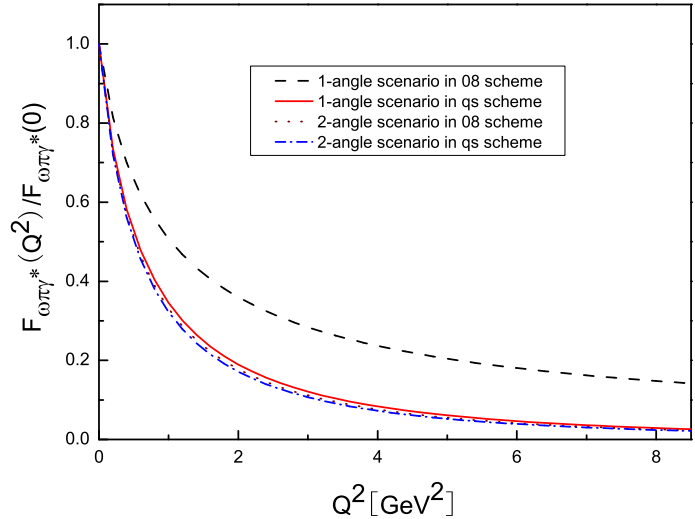}
\caption{\label{fig1} Theoretical prediction of the $Q^2$ behavior
of the normalized form factor
$F_{\omega\rightarrow\pi\gamma^*}(Q^2)/F_{\omega\rightarrow\pi\gamma^*}(0)$
in one-mixing-angle scenario and two-mixing-angle scenario in the
octet-singlet mixing scheme and the quark flavor mixing scheme.}
\end{figure}

\begin{figure}[h]
\centering
\includegraphics[0,0][300,250]{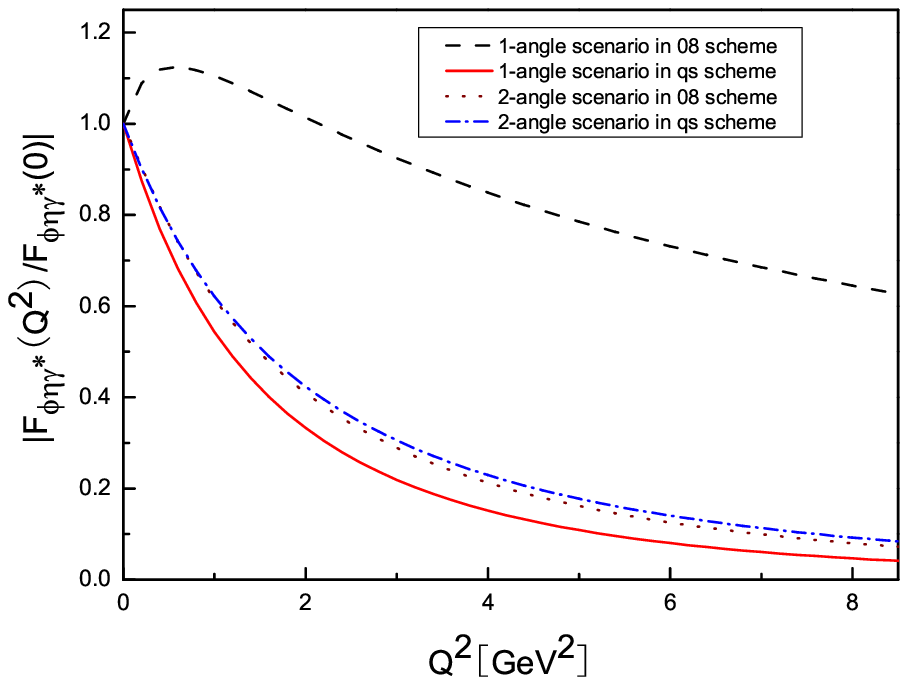}
\caption{\label{fig2} Theoretical prediction of the $Q^2$ behavior
of the normalized form factor
$F_{\phi\rightarrow\eta\gamma^*}(Q^2)/F_{\phi\rightarrow\eta\gamma^*}(0)$
in one-mixing-angle scenario and two-mixing-angle scenario in the
octet-singlet mixing scheme and the quark flavor mixing scheme.}
\end{figure}

\begin{figure}[h]
\centering
\includegraphics[0,0][300,250]{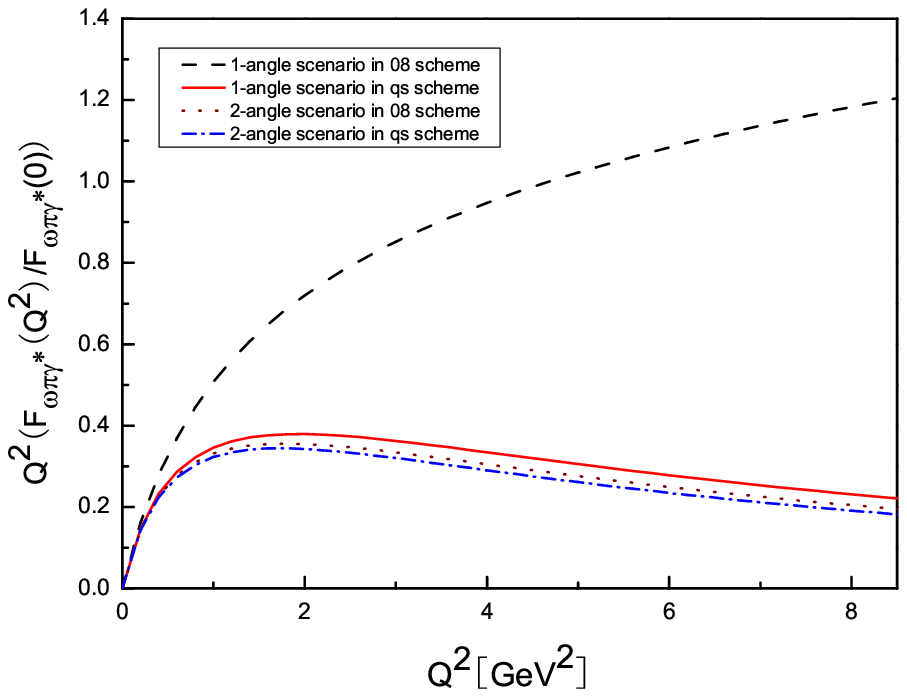}
\caption{\label{fig3} Theoretical prediction of the $Q^2$ behavior
of $Q^2
F_{\omega\rightarrow\pi\gamma^*}(Q^2)/F_{\omega\rightarrow\pi\gamma^*}(0)$
in one-mixing-angle scenario and two-mixing-angle scenario in the
octet-singlet mixing scheme and the quark flavor mixing scheme.}
\end{figure}

\begin{figure}[h]
\centering
\includegraphics[0,0][300,250]{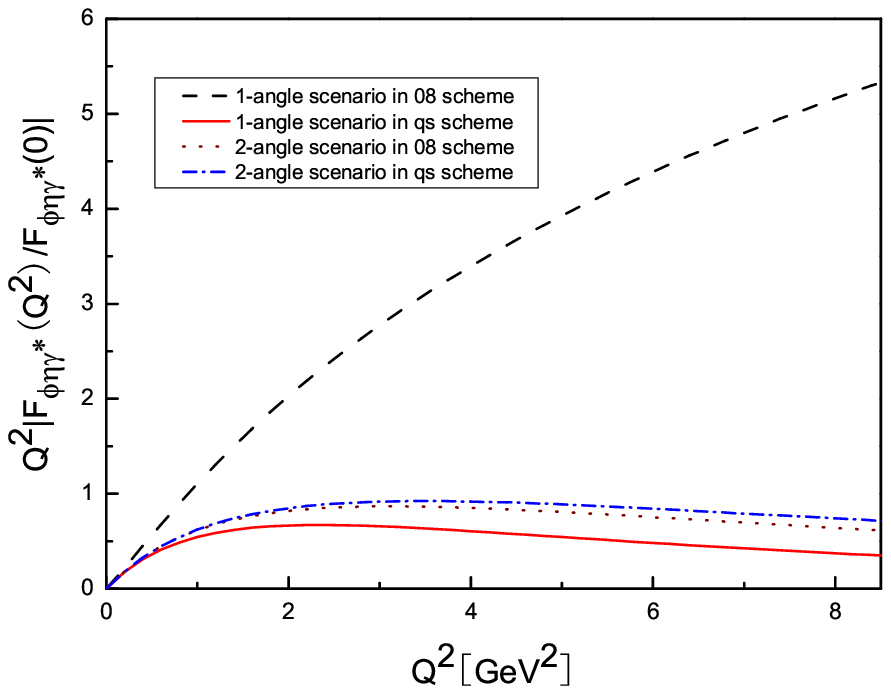}
\caption{\label{fig4} Theoretical prediction of the $Q^2$ behavior
of $Q^2
F_{\phi\rightarrow\eta\gamma^*}(Q^2)/F_{\phi\rightarrow\eta\gamma^*}(0)$
in one-mixing-angle scenario and two-mixing-angle scenario in the
octet-singlet mixing scheme and the quark flavor mixing scheme.}
\end{figure}

With the parameters of $\eta,\eta'$ set, the parameters and mixing
angle of $\omega$-$\phi$ are set together by the decay widths of
$\omega,\phi\rightarrow e^+ e^-$ and the decay widths between
$\omega,\phi$ and $\eta,\eta'$, i.e.  the following constraints:
\begin{eqnarray}
\Gamma_{V\rightarrow e^+ e^-} &=&\frac{4\pi\alpha^2
    f_V^2}{3 m_V}~(V=\omega,\phi), \label{eq:vconstraints1}\\
\Gamma_{V\rightarrow S\gamma}
    &=&\frac{\alpha}{3}
    \left|F_{V\rightarrow S\gamma^*}(0)\right|^2
    \left(\frac{m_V^2-m_S^2}{2 m_V}\right)^3~(V=\omega,\phi; S=\pi,\eta,\eta'),\label{eq:vconstraints2}\\
\Gamma_{S\rightarrow V\gamma}
    &=&\alpha\left|F_{S\rightarrow V\gamma^*}(0)\right|^2
    \left(\frac{m_S^2-m_V^2}{2 m_S}\right)^3 ~(V=\omega;S=\eta').\label{eq:vconstraints3}
\end{eqnarray}
Combining these experimental constraints with
Eqs.~(\ref{eq:fphi}-\ref{eq:Fomega0eta0}), we can get the mixing
angle and parameters of $\phi$ and $\omega$ as listed in the second
column of Table~\ref{tab:table4}. With all the parameters set as
shown in Table~\ref{tab:table3} and Table~\ref{tab:table4}, we can
calculate the $Q^2$ evolving behavior of transition form factors in
the spacelike region according to Eqs.~(\ref{eqn:0-8
Vpig}-\ref{eqn:0-8 Vetag}) as shown in
Fig.~\ref{fig1}-Fig.~\ref{fig4}. Though many efforts were devoted to
determining the branching ratios of the decays $V\rightarrow
P\gamma$ or $P\rightarrow V\gamma$ (with $V=\omega,\phi$;
$P=\pi,\eta,\eta'$), there are no experimental data about their form
factors in spacelike region. However, there are some data about
these form factors in the timelike region obtained through the study
of conversion decays of $V\rightarrow P e^+
e^-$~\cite{Dzhelyadin79,Dzhelyadin81,Kaptari07}. Supposing analytic
continuation of the spacelike transition form factors in our model
in the timelike region according to Ref.~\cite{Choi01}, we get the
timelike transition form factors and compare them with the
experimental data.

\begin{figure}[h]
\centering
\includegraphics[0,0][300,250]{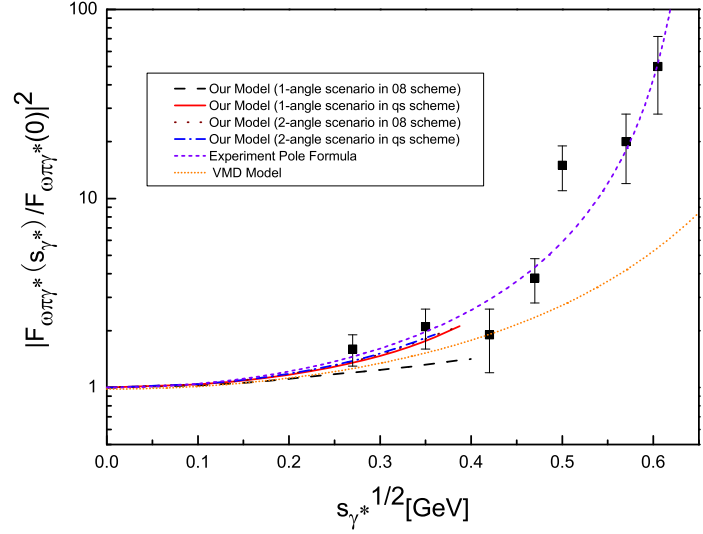}
\caption{\label{fig5}  The $Q^2$ behavior of the normalized form
factor
$F_{\omega\rightarrow\pi\gamma^*}(Q^2)/F_{\omega\rightarrow\pi\gamma^*}(0)$
using one-mixing-angle scenario and two-mixing-angle scenario in the
octet-singlet mixing scheme and the quark flavor mixing scheme
compared with the experimental data~\cite{Kaptari07,Landsberg85} and
the vector meson dominance (VMD) model result in the timelike
region.}
\end{figure}

\begin{figure}[h]
\centering
\includegraphics[0,0][300,250]{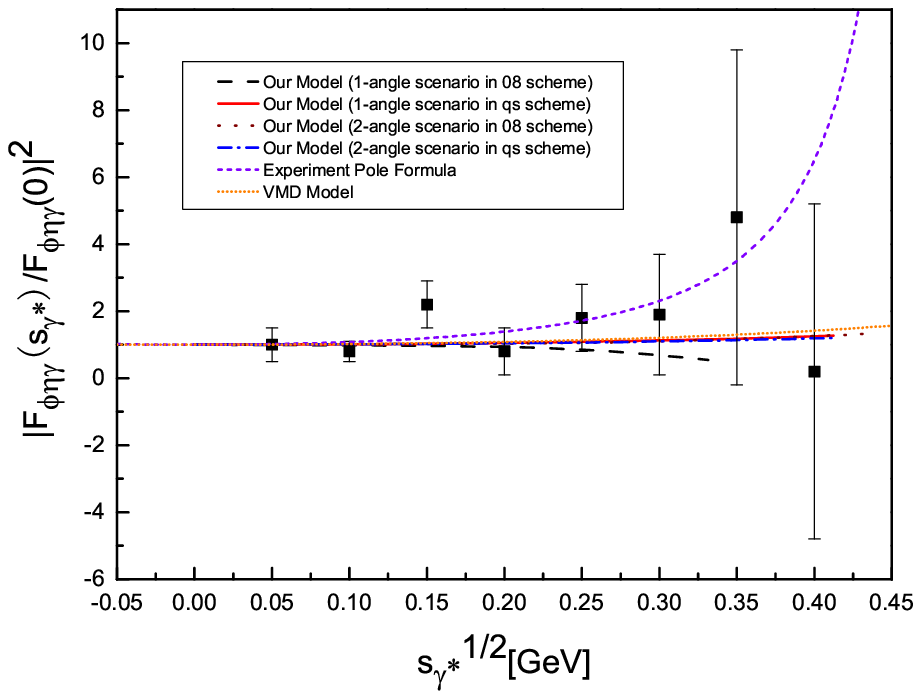}
\caption{\label{fig6}  The $Q^2$ behavior of the normalized form
factor
$F_{\phi\rightarrow\eta\gamma^*}(Q^2)/F_{\phi\rightarrow\eta\gamma^*}(0)$
using one-mixing-angle scenario and two-mixing-angle scenario in the
octet-singlet mixing scheme and the quark flavor mixing scheme
compared with the experimental data~\cite{Achasov01} and the VMD
model result in the timelike region.}
\end{figure}

When changing to the quark flavor mixing scheme, we just make the
replacements $c_8\rightarrow c_q$, $c_0\rightarrow c_s$, while
\begin{eqnarray}
(c_\pi,c_q,c_s)=(1,\frac{5}{3},\frac{\sqrt{2}}{3}).
\end{eqnarray}
We just suppose $\beta_{\eta_q}=\beta_{\eta_s}$ to simplify the
situation. The parameters we get in the quark flavor mixing scheme
are listed in the second columns of Table~\ref{tab:table3} and
Table~\ref{tab:table4}.

\begin{table}
\caption{\label{tab:table3}Experimental values~\cite{PDG08} of the
$\eta$, $\eta'$ decay widths  are compared with theoretical values.
Parameters set in different schemes are listed below.}
\begin{ruledtabular}
\begin{tabular}{cccccc}
     & $F_{\mathrm{exp}}$(GeV)
     & $\begin{array}{c}
        F_{\mathrm{th}}~\mathrm{(GeV)} \\($one-angle scenario$ \\ $in 08 scheme$)
         \end{array}$
     & $\begin{array}{c}
        F_{\mathrm{th}}~\mathrm{(GeV)} \\($one-angle scenario$ \\ $in qs scheme$)
         \end{array}$
     & $\begin{array}{c}
        F_{\mathrm{th}}~\mathrm{(GeV)} \\($two-angle scenario$ \\ $in 08 scheme$)
         \end{array}$
     & $\begin{array}{c}
        F_{\mathrm{th}}~\mathrm{(GeV)} \\($two-angle scenario$ \\ $in qs scheme$)
         \end{array}$\\
  \hline
  $F_{\eta\rightarrow\gamma\gamma^*}(0)$  & $0.272\pm0.007$    & 0.272  & 0.290 & 0.272 & 0.259\\
  $F_{\eta'\rightarrow\gamma\gamma^*}(0)$ & $0.342\pm0.006$  & 0.342   & 0.283 & 0.342 & 0.317\\
  $F_{\rho\rightarrow\eta\gamma^*}(0)$    & $1.59\pm0.05$    &  1.53   &  1.69 & 1.59 & 1.66 \\
  $F_{\eta'\rightarrow\rho\gamma^*}(0)$    & $1.35\pm0.06 $   & 1.74   & 1.34 & 1.35 & 1.42\\
  \hline
  $\theta^S$    &     &$-16.05^\circ $ & $38.29^\circ $
& $\begin{array}{c}
        \theta^S_8=-26.18^\circ \\ \theta^S_0=-2.85^\circ
   \end{array}$
& $\begin{array}{c}
        \theta^S_q=40.57^\circ \\ \theta^S_s=43.89^\circ
   \end{array}$\\
\hline
 Parameters & & $\begin{array}{c}
        A_{\eta8}=27.54 ~\mathrm{GeV}^{-1}\\
        \beta_{\eta8}=0.505 ~\mathrm{GeV}\\
        A_{\eta0}=42.50~\mathrm{GeV}^{-1} \\
        \beta_{\eta0}=0.486 ~\mathrm{GeV}
   \end{array}$
& $\begin{array}{c}
    A_{\eta q}=34.023~\mathrm{GeV}^{-1}\\
    \beta_{\eta q}=0.525~\mathrm{GeV}\\
    A_{\eta s}=54.11~ \mathrm{GeV}^{-1}\\
    \beta_{\eta s}=0.525~ \mathrm{GeV}
\end{array}$
& $\begin{array}{c}
        A_{\eta8}=41.65~ \mathrm{GeV}^{-1}\\
        \beta_{\eta8}=0.607~ \mathrm{GeV}\\
        A_{\eta0}=32.12~\mathrm{GeV}^{-1} \\
        \beta_{\eta0}=0.925~ \mathrm{GeV}
   \end{array}$
& $\begin{array}{c}
    A_{\eta q}=34.40~\mathrm{GeV}^{-1}\\
    \beta_{\eta q}=0.525~ \mathrm{GeV}\\
    A_{\eta s}=91.39~ \mathrm{GeV}^{-1}\\
    \beta_{\eta s}=0.525~ \mathrm{GeV}
\end{array}$
\end{tabular}
\end{ruledtabular}
\end{table}

\begin{table}
\caption{\label{tab:table4}Experimental data~\cite{PDG08} for the
decay constants and decay widths of $\omega,\phi$ are compared with
theoretical values. Parameters set in different schemes are listed
below.}
\begin{ruledtabular}
\begin{tabular}{cccccc}
    & $F_{\mathrm{exp}}/f_{\mathrm{exp}}~(\mathrm{GeV})$
     & $\begin{array}{c}
        F_{\mathrm{th}}/f_{\mathrm{th}}~(\mathrm{GeV}) \\($one-angle scenario$ \\ $in 08 scheme$)
         \end{array}$
     & $\begin{array}{c}
        F_{\mathrm{th}}/f_{\mathrm{th}}~(\mathrm{GeV}) \\($one-angle scenario$ \\ $in qs scheme$)
         \end{array}$
     & $\begin{array}{c}
        F_{\mathrm{th}}/f_{\mathrm{th}}~(\mathrm{GeV}) \\($two-angle scenario$ \\ $in 08 scheme$)
         \end{array}$
     & $\begin{array}{c}
        F_{\mathrm{th}}/f_{\mathrm{th}}~(\mathrm{GeV}) \\($two-angle scenario$ \\ $in qs scheme$)
         \end{array}$\\
  \hline
  $f_\phi(\phi\rightarrow e^+ e^-)$  & $0.076\pm 0.012$ & 0.076  & 0.076 & 0.068 & 0.076\\
  $f_\omega(\omega\rightarrow e^+ e^-)$ & $0.0459\pm0.0008$ & 0.0458 & 0.0459 & 0.0475 & 0.0456\\
  $F_{\phi\rightarrow\pi\gamma^*}(0)$    & $0.133\pm0.003$  &  0.133 &  0.133 & 0.131& 0.132\\
  $F_{\omega\rightarrow\pi\gamma^*}(0)$    & $2.385\pm0.004 $ & 2.080  & 2.385 & 2.327 & 2.295\\
  $F_{\phi\rightarrow\eta\gamma^*}(0)$    & $-0.692\pm0.007 $  & $-0.135$ & $-0.573$ & $-0.581$ & $-0.662$\\
  $F_{\phi\rightarrow\eta'\gamma^*}(0)$    & $0.712\pm0.01$     & 0.267  & 0.787 & 0.853 & 0.742\\
  $F_{\omega\rightarrow\eta\gamma^*}(0)$    & $0.449\pm0.02$  & 0.477  & 0.572 & 0.453 & 0.457\\
  $F_{\eta'\rightarrow\omega\gamma^*}(0)$   & $0.460\pm0.03$  & 0.482 & 0.383 & 0.450 & 0.470\\
\hline
  $\theta^V$  &  & $42.20^\circ $ &$86.82^\circ$
& $\begin{array}{c}
        \theta^V_8=12.17^\circ \\ \theta^V_0=77.82^\circ
         \end{array}$
& $\begin{array}{c}
        \theta^V_q=86.71^\circ \\ \theta^V_s=93.43^\circ
         \end{array}$\\
\hline
Parameters & & $\begin{array}{c}
    A_{\omega8}=39.78~\mathrm{ GeV}^{-1}\\
    \beta_{\omega8}=0.481~ \mathrm{GeV}\\
    A_{\omega0}=17.58~ \mathrm{GeV}^{-1}\\
    \beta_{\eta0}=3.726~\mathrm{ GeV}
\end{array}$
&$\begin{array}{c}
    A_{\omega q}=46.15~ \mathrm{GeV}^{-1}\\
    \beta_{\omega q}=0.374~ \mathrm{GeV}\\
    A_{\omega s}=579.96~ \mathrm{GeV}^{-1}\\
    \beta_{\omega s}=0.291~ \mathrm{GeV}
\end{array}$
& $\begin{array}{c}
    A_{\omega8}=215.18~ \mathrm{GeV}^{-1}\\
    \beta_{\omega8}=0.332~ \mathrm{GeV}\\
    A_{\omega0}=135.52~ \mathrm{GeV}^{-1}\\
    \beta_{\eta0}=0.358~ \mathrm{GeV}
\end{array}$
&$\begin{array}{c}
    A_{\omega q}=51.58~ \mathrm{GeV}^{-1}\\
    \beta_{\omega q}=0.330~ \mathrm{GeV}\\
    A_{\omega s}=52.28~ \mathrm{GeV}^{-1}\\
    \beta_{\omega s}=0.490~ \mathrm{GeV}
\end{array}$
\end{tabular}
\end{ruledtabular}
\end{table}

From Table~\ref{tab:table3} and Table~\ref{tab:table4} we can see
that the results in the quark flavor scheme are better than those in
the octet-singlet scheme in reproducing the decay widths related to
the pseudoscalar and vector meson mixing. Concerning their $Q^2$
behaviors after normalized by $F(Q^2)/F(0)$, the results of two
schemes can be compared with each other as shown in
Fig.~\ref{fig1}-Fig.~\ref{fig4}. Extrapolating $Q^2$ to the timelike
region by $q_\perp\rightarrow i q_\perp $~\cite{Choi01}, we get form
factors in timelike $Q^2$ region compared with the experimental data
in Fig.~\ref{fig5} and Fig.~\ref{fig6}. But this region is limited
due to the appearance of a singularity in the numerical calculation;
i.e., the form factors in a large range of time-like region cannot
be calculated simply through analytic extrapolation. In the limited
timelike region our results are comparable with the experimental
data. In the process $\omega\rightarrow\pi\gamma^*$, the timelike
transition form factor produced by the quark flavor scheme is closer
to the experimental pole formula simulation comparing to the
octet-singlet scheme and the vector meson dominance models as shown
in Fig.~\ref{fig5} and Fig.~\ref{fig6}. In the process
$\phi\rightarrow\eta\gamma^*$, the timelike transition form factors
produced by the model are comparable to the data while there are big
error bars in the experimental data. More experimental data are
needed to reduce the error bars.

The mixing angles we get in two schemes are, respectively,
$\theta^S_{08}=-16.05^\circ$, $\theta^V_{08}=42.20^\circ$,
$\theta^S_{qs}=38.29^\circ$ and $\theta^V_{qs}=86.82^\circ$. The
pseudoscalar mixing angles approximately follow the ideal SU(3)
relation $\theta^S_{qs}=\theta^S_{08}+54.7^\circ$. They are
comparable with the mixing angles determined by mass relation in PDG
(2008) and other papers~\cite{Escribano05,Choi99}. But the vector
mixing angles do not follow the relationship
$\theta^V_{qs}=\theta^V_{08}+54.7^\circ$. If we replace
$\theta^V_{qs}-90^\circ\rightarrow \tilde{\theta}^V_{qs}$, i.e.,
change to another mixing expression:
\begin{eqnarray}
\left(
  \begin{array}{c}
    \phi \\
    \omega\\
  \end{array}
\right) = \left(
  \begin{array}{cc}
    \cos\tilde{\theta}^V_{qs} & -\sin\tilde{\theta}^V_{qs} \\
    \sin\tilde{\theta}^V_{qs} & \cos\tilde{\theta}^V_{qs} \\
  \end{array}
\right) \left(
  \begin{array}{c}
    -s\bar{s}~\varphi^s \\
    \frac{1}{\sqrt{2}}(u\bar{u}+d\bar{d})~\varphi^q\\
  \end{array}
\right),
\end{eqnarray}
it can be seen that the mixing angle from the quark flavor mixing
scheme $\tilde{\theta}^V_{qs}=-3.18^\circ$ has the opposite sign
compared with one from the octet-singlet mixing scheme
$\theta^V_{08}+54.7^\circ-90^\circ=6.9^\circ$ [suppose they can be
compared through the ideal SU(3) relationship
$\theta_{qs}=\theta_{08}+54.7^\circ$], and their absolute values are
comparable with each other and also with the vector meson mixing
angle coming from the mass relation in PDG (2008). In other papers,
sometimes the vector meson mixing angle is
positive~\cite{Bramon97,Escribano05,Ohshima80} and sometimes it is
negative~\cite{Benayoun99,Benayoun00,Kucurkarslan06}, and there is
always an alternative phase convention~\cite{Scadron84}. However,
their absolute values are all comparable with each other
approximately. In this paper we do not use the other phase
convention, but just adopt the real rotation of the octet-singlet or
quark flavor bases. The results seem to prefer the quark flavor
mixing scheme, which gives us a negative vector mixing angle, when
introducing only one mixing angle.

The reproduction of decay widths with the one-mixing-angle scenario
in the octet-singlet mixing scheme is not as good as that in the
quark flavor mixing scheme. In order to improve the octet-singlet
mixing scheme, it is natural to introduce two-mixing-angle scenario.

\subsection{\label{sec:level4.3}Set $\eta\eta'$, $\phi\omega$
parameters in the two-mixing-angle scenario in two mixing schemes}
The two-mixing-angle scenario was introduced to study pseudoscalar
meson $\eta$-$\eta'$ mixing, especially concerning the decay
constants~\cite{Leutwyler98,Kaiser98,Feldmann98}. Here we try to
introduce two mixing angles to study the vector meson
$\omega$-$\phi$ mixing.

First we restudy $\eta$-$\eta'$ mixing with two-mixing-angle
scenario in the octet-singlet mixing scheme. When introducing two
mixing angles, we cannot have explicit solution of $\theta^S_0$,
$\theta^S_8$ like in the one-mixing-angle scenario in
Eqs.~(\ref{eq:mixingangleS}). Using Eqs.~(\ref{eq:Q2infty1},
\ref{eq:Q2infty2}, \ref{eq:PCAC2angle1}-\ref{eq:etap2g}) as
constraints, we can set the pseudoscalar meson mixing angles and the
parameters of $\eta,\eta'$. The reproduction of experimental data
can be improved as shown in Table~\ref{tab:table3}.

 With the parameters of $\eta,\eta'$ set, we can proceed to set the
parameters of $\omega,\phi$ in the two-mixing-angle scenario in
octet-singlet mixing scheme. Using constraints
Eqs.~(\ref{eq:vconstraints1}-\ref{eq:vconstraints3}) combined with
Eqs.~(\ref{eq:fphi}-\ref{eq:Fomega0eta0}), we get the parameters and
reproduction of the decay widths listed in the fourth column in
Table~\ref{tab:table4}. Obviously the experimental data are better
reproduced in the two-mixing-angle scenario than in the
one-mixing-angle scenario.

Though the two mixing angles we get deviate a lot from each other(
$\Delta\theta^S_{08}=\theta^S_0-\theta^S_8=23.33^\circ$,
$\Delta\theta^V_{08}=\theta^V_0-\theta^V_8=65.65^\circ$), the
average value of the two mixing angles is comparable with the
one-mixing angle result:
$\overline{\theta^S_{08}}=\frac{\theta^S_0+\theta^S_8}{2}=-14.52^\circ\sim~\theta^S_{08}=-16.05^\circ$,
$\overline{\theta^V_{08}}=\frac{\theta^V_0+\theta^V_8}{2}=45.00^\circ\sim~\theta^V_{08}=42.20^\circ$.

As reviewed in Ref.~\cite{Feldmann00}, both octet-singlet mixing
scheme and quark flavor mixing scheme can be introduced with two
mixing angles, while the results from $\eta$-$\eta'$ study show that
the difference of the two mixing angles in the quark flavor scheme
is much smaller than that in the octet-singlet scheme. This suggests
that the one-mixing-angle approximation is more reasonable in the
quark flavor scheme. So we introduce two-mixing-angle scenario in
the quark flavor mixing scheme to study not only $\eta$-$\eta'$
mixing but also $\omega$-$\phi$ mixing. The steps are similar to
those in the octet-singlet mixing scheme, just by changing the octet
and singlet bases to the quark flavor bases and replacing the
constants $c_8,c_0$ by $c_q,c_s$. The results we get are listed in
the final columns of Tables~\ref{tab:table3} and
Table~\ref{tab:table4}.

We can see that the reproduction of the experimental data in the
two-mixing-angle scenario is also improved compared with the
one-mixing-angle scenario in the quark flavor scheme. The
differences between the two mixing angles in the quark flavor scheme
are much smaller than those in the octet-singlet scheme:
$\Delta\theta^S_{qs}=\theta^S_s-\theta^S_q=3.32^\circ\ll\Delta\theta^S_{08}$,
$\Delta\theta^V_{qs}=\theta^V_s-\theta^V_q=6.72^\circ\ll\Delta\theta^V_{08}$.
And their average values are also close to the one-mixing-angle
scenario results:
$\overline{\theta^S_{qs}}=\frac{\theta^S_s+\theta^S_q}{2}=42.23^\circ\sim~\theta^S_{qs}=38.29^\circ$,
$\overline{\theta^V_{qs}}=\frac{\theta^V_s+\theta^V_q}{2}=90.07^\circ\sim~\theta^V_{qs}=86.82^\circ$.
These results also explain why one-mixing-angle approximation in the
quark flavor scheme is more reasonable when studying $\eta$-$\eta'$
and $\omega$-$\phi$ mixing.

Now we have four set of parameters which can be used to reproduce
the decay widths of the mesons and calculate the transition form
factors of the mesons. The reproduction of the decay widths is
improved by introducing two mixing angles in both schemes. The $Q^2$
evolving behavior of the transition form factors is shown and
compared in Fig.~\ref{fig1}-Fig.~\ref{fig6}. It is interesting to
notice that the three curves we get from the one-mixing-angle
scenario in the quark flavor scheme, and the two-mixing-angle
scenario in the quark flavor scheme and octet-singlet scheme are
close to each other; however, the curve produced by the
one-mixing-angle scenario in the ocetet-singlet scheme deviates from
the other three. Concerning the $Q^2$ behavior of the meson form
factors, the mixing-angle results and the decay widths fit by the
model, the best choices is the two-mixing-angle scenario in the
quark flavor mixing scheme and in the octet-singlet mixing scheme.
The one-mixing-angle scenario in the quark flavor scheme is also
acceptable, while the one-angle-mixing scenario in the octet-singlet
mixing scheme deviates a lot from the other three and may be the
worst one of the four choices.

\section{Conclusion}
The light-cone quark model is a useful approach to study hadronic
properties in low energy region which is related to nonperturbative
QCD. With the decay widths, form factors and radii of the mesons as
constraints, we set the mixing angles and wave function parameters
of the pseudoscalar mesons $\eta,\eta'$ and the vector mesons
$\omega,\phi$ with two mixing angle scenarios in two mixing schemes.
Comparing theoretical results with experimental data, we find that
the results from the quark flavor mixing scheme are better than
those from the octet-singlet mixing scheme and the results of the
two-mixing-angle scenario are better than those of the
one-mixing-angle scenario.  We calculate the transition form factors
in the spacelike region using the two mixing angle scenarios in two
mixing schemes,  respectively, and compare their behavior. By
extrapolating the form factors to the limited timelike region, our
results are comparable with the experimental data. The absolute
values of vector meson mixing angles we get in two mixing schemes
are comparable with each other. If one only introduces one mixing
angle to study processes related to pseudoscalar and vector meson
mixing, the quark flavor mixing scheme is more reliable than the
octet-singlet mixing scheme. When introducing two mixing angles,
both schemes work well.

\section*{Ackonwledgments}

This work is partially supported by National Natural Science
Foundation of China (Nos.~10721063, 10575003, 10528510), by the Key
Grant Project of Chinese Ministry of Education (No.~305001), by the
Research Fund for the Doctoral Program of Higher Education (China).

\appendix

\section{\label{app:a}}
After getting the wave functions of the mesons through the
Melosh-Wigner rotation or vertices in Eq.~(\ref{wavefucntion})
equivalently, we can calculate the decay constant $f_P$ of a charged
pseudoscalar meson P:
\begin{equation}
f_P=I_{P\mu\nu}[m_{q_1},m_{q_2},A_P,\beta_P],
\end{equation}
in which
 \begin{eqnarray}
I_{P\mu\nu}[m_{q_1},m_{q_2},A_P,\beta_P] =2\sqrt{3}
 \int \frac{\mathrm{d}x \mathrm{d}^2\mathbf{k}_\perp}{16\pi^3}
~\varphi_P(\mathbf{k}_\perp)\frac{m_{q_1}(1-x)+m_{q_2}x}{\sqrt{\mathbf{k}_\perp^2+(m_{q_1}(1-x)+m_{q_2}x)^2}}
.
\end{eqnarray}

The form factor of a pseudoscalar meson P is
\begin{eqnarray}
F_P(Q^2)=Q_{q_1} I_{PP}[m_{q_1},m_{q_2},A_P,\beta_P]+Q_{q_2}
I_{PP}[m_{q_2},m_{q_1},A_P,\beta_P],
\end{eqnarray}
in which
\begin{eqnarray}
I_{PP}[m_{q_1},m_{q_2},A_P,\beta_P]
    &=& \int \frac{\mathrm{d}x \mathrm{d}^2\mathbf{k}_\perp}{16\pi^3}
         ~\varphi_P^*(x,\mathbf{k}'_\perp)\varphi_P(x,\mathbf{k}_\perp)\nonumber\\
    &&\times \frac{(m_{q_1}(1-x)+m_{q_2} x)^2+\mathbf{k}_\perp\cdot \mathbf{k}'_\perp}
         {\sqrt{(m_{q_1}(1-x)+m_{q_2} x)^2+\mathbf{k}_\perp^2}\sqrt{(m_{q_1}(1-x)+m_{q_2} x)^2+\mathbf{k}'^2_\perp}}.
\end{eqnarray}

The transition form factor of a pseudoscalar meson
$F_{P\rightarrow\gamma\gamma^*}(Q^2)$ is
\begin{eqnarray}
F_{P\rightarrow\gamma\gamma^*}(Q^2) = Q_q^2
        ~I_{P\gamma\gamma^*}[m_q,A_P,\beta_P],
\end{eqnarray}
in which
\begin{eqnarray}
I_{P\gamma\gamma^*}[m_q,A_P,\beta_P]
    &=& 4\sqrt{6}
    \int \frac{\mathrm{d}x \mathrm{d}^2\mathbf{k}_\perp}{16\pi^3}~\varphi_P(x,\mathbf{k}_\perp)
    \frac{m_q}{x\sqrt{\mathbf{k}_\perp^2+m^2}}\frac{x(1-x)}{m_q^2+\mathbf{k}_\perp^{'2}}.
\end{eqnarray}

The decay constant of a neutral vector meson V is
\begin{equation}
f_V =2\sqrt{3}\int \frac{\mathrm{d}x
\mathrm{d}^2\mathbf{k}_\perp}{16\pi^3}
  \frac{1}{\sqrt{x(1-x)}}
  ~\varphi_V(x,\mathbf{k}_\perp)
  \frac{2\mathbf{k}_\perp^2+m_q(\mathcal{M}+2m_q)}{\sqrt{\mathbf{k}_\perp^2+m_q^2}(\mathcal{M}+2m_q)}.
\end{equation}

\section{\label{app:b}}
 In two-mixing-angle scenario in the quark flavor mixing scheme, the mixing of the vector mesons is defined
 by
 \begin{eqnarray}
\left(\begin{array}{c}
         |\phi\rangle \\
        |\omega\rangle
    \end{array}\right)
&=&\left(\begin{array}{cc}
     \cos \theta^V_{q} & -\sin \theta^V_{s} \\
     \sin \theta^V_{q} & \cos \theta^V_{s} \\
   \end{array}\right)
\left(\begin{array}{c}
    |\omega_q\rangle \\
    |\omega_s\rangle \\
   \end{array}\right);
\end{eqnarray}
the decay constants and transition form factors of the vector mesons
are:
\begin{eqnarray}
\left(
  \begin{array}{c}
    f_\phi \\
    f_\omega\\
  \end{array}
\right) = \left(
  \begin{array}{cc}
    \cos\theta^V_{q} & -\sin\theta^V_{s} \\
    \sin\theta^V_{q} & \cos\theta^V_{s} \\
  \end{array}
\right) \left(
  \begin{array}{c}
    f_{\omega_q} \\
    f_{\omega_s} \\
  \end{array}
\right),
\end{eqnarray}

\begin{eqnarray}
\left(
  \begin{array}{c}
    F_{\phi\rightarrow\pi\gamma^*} (Q^2)\\
    F_{\omega\rightarrow\pi\gamma^*}(Q^2)\\
  \end{array}
\right) = \left(
  \begin{array}{cc}
    \cos\theta^V_{q} & -\sin\theta^V_{s} \\
    \sin\theta^V_{q} & \cos\theta^V_{s} \\
  \end{array}
\right) \left(
  \begin{array}{c}
    F_{\omega_q\rightarrow\pi\gamma^*}(Q^2) \\
    0 \\
  \end{array}
\right), \label{eqn:q-s Vpig}
\end{eqnarray}

\begin{eqnarray}
\left(
  \begin{array}{c}
    F_{\phi\rightarrow\eta\gamma^*} (Q^2)\\
    F_{\phi\rightarrow\eta'\gamma^*} (Q^2)\\
    F_{\omega\rightarrow\eta\gamma^*}(Q^2)\\
    F_{\eta'\rightarrow\omega\gamma^*}(Q^2)
  \end{array}
\right) = \left(
  \begin{array}{cc}
    \cos\theta^V_{q} & -\sin\theta^V_{s} \\
    \sin\theta^V_{q} & \cos\theta^V_{s} \\
  \end{array}
\right) \otimes \left(
  \begin{array}{cc}
    \cos\theta^S_{q} & -\sin\theta^S_{s} \\
    \sin\theta^S_{q} & \cos\theta^S_{s} \\
  \end{array}
\right) \left(
  \begin{array}{c}
    F_{\omega_q\rightarrow\eta_q\gamma^*}(Q^2) \\
    0 \\
    0\\
    F_{\omega_s\rightarrow\eta_s\gamma^*}(Q^2) \\
  \end{array}
\right), \label{eqn:q-s Vetag}
\end{eqnarray}
in which,
\begin{equation}
\left\{
\begin{array}{lll}
 F_{\omega_q\rightarrow\pi\gamma^*}(Q^2) &=&
    \frac{1}{\sqrt{2}}\frac{1}{\sqrt{2}} (2Q_u
    I_{VP\gamma}[m_q,A_{\omega_q},\beta_{\omega_q},A_\pi,\beta_\pi]-2 Q_d
    I_{VP\gamma}[m_q,A_{\omega_q},\beta_{\omega_q},A_\pi,\beta_\pi])\\
&=&
    I_{VP\gamma}[m_q,A_{\omega_q},\beta_{\omega_q},A_\pi,\beta_\pi]\\
F_{\omega_q\rightarrow\eta_q\gamma^*}(Q^2) &=&
    \frac{1}{\sqrt{2}}\frac{1}{\sqrt{2}} (2Q_u
    I_{VP\gamma}[m_q,A_{\omega_q},\beta_{\omega_q},A_{\eta_q},\beta_{\eta_q}]+2 Q_d
    I_{VP\gamma}[m_q,A_{\omega_q},\beta_{\omega_q},A_{\eta_q},\beta_{\eta_q}])\\
&=&
    \frac{1}{3}I_{VP\gamma}[m_q,A_{\omega_q},\beta_{\omega_q},A_{\eta_q},\beta_{\eta_q}]\\
F_{\omega_s\rightarrow\eta_s\gamma^*}(Q^2) &=&
    2Q_S
    I_{VP\gamma}[m_S,A_{\omega_s},\beta_{\omega_s},A_{\eta_s},\beta_{\eta_s}].\\
\end{array}
\right.
\end{equation}

In the two-mixing-angle scenario, $\theta^S_q$, $\theta^S_s$,
$\theta^V_q$ and $\theta^V_s$ are fit separately. When setting
$\theta^S_q=\theta^S_s=\theta^S_{qs}$,
$\theta^V_q=\theta^V_s=\theta^V_{qs}$, one returns back to the
one-mixing-angle scenario.

\end{document}